\documentclass{aastex63}

\received{June 1, 2019}
\revised{January 10, 2019}
\accepted{\today}
\submitjournal{ApJL}

\shorttitle{Sample article}
\shortauthors{Zaqarashvili}

\begin{document}

\title{Dynamic kink instability and transverse motions of solar spicules}

\correspondingauthor{Teimuraz Zaqarashvili}
\email{teimuraz.zaqarashvili@uni-graz.at}

%
%

\author{Teimuraz V. Zaqarashvili}
\affiliation{IGAM, Institute f\"ur Physik, University of Graz, Universit\"atsplatz 5, 8010 Graz, Austria}
\affiliation{Ilia State University, Cholokashvili ave 5/3, Tbilisi, Georgia}
\affiliation{Abastumani Astrophysical Observatory, Mount Kanobili, Abastumani, Georgia}

\nocollaboration{1}

%
%
%
%



\begin{abstract}

Hydrodynamic jets are unstable to the kink instability ($m=1$ mode in cylindrical geometry) owing to the centripetal force, which increases the transverse displacement of the jet. When the jet moves along a magnetic field, then the Lorentz force tries to decrease the displacement and stabilises the instability of sub-Alfv\'enic flows. The threshold of the instability depends on the  Alfv\'en Mach number (the ratio of Alfv\'en and jet speeds). We suggest that the dynamic kink instability may be of importance to explain observed transverse motions of type II spicules in the solar atmosphere. We show that the instability may start for spicules which rise up at the peripheries of vertically expanding magnetic flux tubes owing to the decrease of the Alfv\'en speed in both, the vertical and the radial directions. Therefore, inclined spicules may be more unstable and have more higher transverse speeds. Periods and growth times of unstable modes in the conditions of type II spicules have the values of $30$ s and $25-100$ s, respectively, which are comparable to the life time of the structures. This may indicate to the interconnection between high speed flow and rapid disappearance of type II spicules in chromospheric spectral lines.

\end{abstract}

\keywords{Sun: atmosphere --
                Sun: oscillations --
                Physical data and processes: Instabilities}


\section{Introduction} \label{sec:intro}

Spicules are dense chromospheric plasma jets flowing upwards into hot and tenuous corona \citep{Beckers1968}. They are usually observed in chromospheric
H$\alpha$, D$_3$ and Ca II H lines at the solar limb. Disc counterparts of spicules are mottles, which have almost the same properties as spicules \citep{Tsiropoula1997}.
Typical life time  and upward velocity of classical spicules/mottles are 5-15 min and $\sim$ 20-30 km s$^{-1}$, respectively. Recent Hinode/SOT observations with high spatial and
temporal resolutions \citep{DePontieu2007a} revealed another type of spicules (type II). \citet{Rouppe2009} reported disk counterparts of type II spicules named as Rapid Blue/Red shifted excursions (RBE/RREs). Both, type II spicules and RBE/RREs have much shorter life time (about 10-150 s) and higher upward velocities (50-150 km s$^{-1}$) than classical (type I) spicules  in Ca II H \citep{DePontieu2007a}, H$\alpha$ \citep{Kuridze2015}, and Ly$\alpha$ \citep{Chintzoglou2018} lines. However, recent studies combing observations from Hinode, VAULT2.0 and  the Interface Region Imaging Spectrograph (IRIS) showed that after disappearance in chromospheric lines many type II spicules appear in hotter lines like Mg II, C II and Si IV \citep{Pereira2014,Rouppe2015,Skogsrud2015,Chintzoglou2018}. 

Observations show that spicules undergo continuous transverse motion of their axes. Type I spicules and mottles show obvious oscillatory transverse motions of axes with periods of 30-500 s interpreted as Alfv\'en and/or magnetohydrodynamic (MHD) kink waves \citep{Kukhianidze2006,DePontieu2007b,Zaqarashvili2007,Zaqarashvili2009,Okamoto2011,Kuridze2012,Tsiropoula2012}. On the other hand, the short life-time of type II spicules and RBE/RREs in chromospheric lines complicates to observe the full swing of the axes, but most of the structures show linear trend of transverse motion \citep{Kuridze2015}: the structures move in transverse direction and do not return back, but disappear. Moreover, the transverse velocity is about 8  km  s$^{-1}$ in mottles \citep{Kuridze2012} and almost twice (14-17 km  s$^{-1}$) in RBE/RREs \citep{Kuridze2015}. Do type II spicules disappear over shorter time than the oscillation period? No existing mechanism supports such strong damping of oscillations and corresponding heating. Even if such strongly damped oscillations exist, then why they are not damped in type I spicules?  On the other hand, appearance of spicules in IRIS spectral lines show that type II spicules are rapidly heated to transition region (TR) temperatures \citep{Pereira2014}, though the heating mechanism is not yet completely clear. Ion-neutral collisions, Kelvin-Helmholtz instability or both together might lead to the rapid heating \citep{Kuridze2016,Martinez2017,Antolin2018}, but it is not yet fully established. If spicules are rapidly heated, then their axes may continue to oscillate in TR lines.  Another possibility is that the spicules are quickly destroyed by some instability process. This point is still open for further discussion.

\citet{DePontieu2012} showed that type II spicules are characterized by the simultaneous action of three different types of motion such as field-aligned flows, swaying and torsional motions, though the field-aligned flows are few time stronger than the others. 
In this letter, we assume spicules as field-aligned jets and propose that dynamic kink instability of jets could explain the observed transverse (swaying) motion of spicule axes. The dynamic instability of jets is well known in nonmagnetic fluid dynamics, where the jets show antisymmetric displacement of axis \citep{Drazin2002}. In cylindrical geometry, axially symmetric nonmagnetic jets have the vorticity component in $\phi$ direction, hence they can be considered as vortically twisted tubes. Therefore, the jets are unstable to the dynamic kink instability very similar to the magnetically twisted tubes, which are unstable to the MHD kink instability. Flow-aligned magnetic field usually stabilises sub-Alfv\'enic jets, but super-Alfv\'enic jets still can be unstable to the dynamic kink instability. This process can be of importance in high-speed type II spicules and RBE/RREs.

\section{Dynamic kink instability in jets} \label{sec:style}

We consider a homogeneous cylindrical jet of uniform velocity $U_z$, radius $a$ and density $\rho_0$, which moves in a medium of uniform density, $\rho_e$, along an uniform magnetic field, $B_z$. This is a very simplified case comparing to the complex structure of the solar chromosphere, though the consideration shows basic properties of the instability. The linear incompressible dynamics of the jet is governed by the dispersion equation (see details in Appendix A)
\begin{equation}\label{1}
(\rho_e E_m-\rho_0)\omega^2+2\rho_0k_z U_z \omega-(\rho_0k^2_z U^2_z+\rho_ek^2_z V^2_{Ae} E_m-\rho_0k^2_zV^2_{A0})=0,
\end{equation}
where $\omega$ is the wave frequency, $V_{Ae}=B_z/\sqrt{4\pi\rho_e}$ and $V_{A0}=B_z/\sqrt{4\pi\rho_0}$ are Alfv\'en speeds outside and inside the jet, while $E_m(k_z a)=(I'_m(k_z a)/I_m(k_z a))(K_m(k_z a)/K'_m(k_z a))$. $I_m(k_z a)$ and $I_m(k_z a)$ are modified Bessel functions with an order $m$ and prime sign $'$ means the differentiation with Bessel function argument. This equation can be derived from the dispersion relation of moving twisted magnetic flux tube (Equation 25 in \citet{Zaqarashvili2014b}) assuming zero twist. Solution of this equation is 
\begin{equation}\label{2}
\omega=\frac{-\rho_0k_zU_z \pm \sqrt{\rho_0 \rho_ek^2_zU^2_zE_m+(\rho_e E_m-\rho_0)(E_m-1)(k^2_z B^2_z/4\pi)} }{\rho_e E_m-\rho_0}.
\end{equation}
Here the complex frequency means the instability of corresponding mode, where the imaginary part of frequency shows the growth rate. In the case of hydrodynamic jet (with $B_z=0$), the anti-symmetric ($m=1$) mode always has imaginary part as $E_1(k_z a)<0$. For long wave length approximation, $k_z a \ll 1$, we have $E_1(k_z a) \approx -1$ and the growth rate is $\omega_i=k_z U_z/2$ for $\rho_e=\rho_0=\rho$. This mode leads to the transverse displacement of the jet axis (see Fig.1) and is called as a kink mode (for a rectangular jet see \citet{Drazin2002}). The similar antisymmetric mode in magnetic tubes is the MHD kink mode \citep{Edwin1983}. However, in static MHD, this mode is stable in non-twisted and weakly twisted tubes, but becomes unstable when the twist exceeds some critical value \citep{Lundquist1951}, hence the mechanism for instability is the Lorentz force. In hydrodynamic jets, the kink mode is unstable for any speed and wavelength. The mechanism of instability is connected to the centripetal force, which acts on the jet moving on the curved trajectory and increases the curvature (Fig. 1). Therefore, the instability can be called as a {\it dynamic kink instability} in difference of MHD kink instability. The jet-aligned magnetic field stabilises the instability as the Lorentz force is a restoring force in this case and tries to decrease the curvature (Fig. 1). This is easily seen from Eq. (2). If we consider the long wavelength limit, then the second term in the square root is $4\rho k^2_z B^2_z/4\pi$. Consequently, the frequency is real when $V_A/U_z > 1/2$, where $V_A=B_z/\sqrt{4\pi \rho}$ is the Alfv\'en speed, therefore the jet is stable. Hence, super-Alfv\'enic jets could be unstable to the kink instability even in non-twisted magnetic configurations. Figure 2 shows the real and imaginary parts of frequency vs longitudinal wave number of kink ($m=1$) mode in dense jets (resembling spicules) for different  Alfv\'en Mach number, i.e. the ratios of Alfv\'en and flow speeds, $V_{Ae}/U_z$, where $V_{Ae}$ is the external Alfv\'en speed. The figure shows that the jet is unstable for all wavelength perturbations when Alfv\'en Mach number equals 0.6 (red line). When $V_{Ae}/U_z=0.7$ then only long wavelength perturbations ($k_z a < 0.1$) are unstable (green line). The jet is completely stable when $V_{Ae}/U_z=0.8$ (blue line). Hence, the dense jet is unstable to the dynamic kink instability when the jet speed is higher than
\begin{equation}\label{3}
U_z>1.25 \, V_{Ae}.
\end{equation}
In the next section we study the instability in typical spicule parameters keeping in mind that the consideration is highly simplified comparing to the real structure of the solar atmosphere.

\section{Kink Instability in spicules}

Spicules are chromospheric jets rising upwards into the corona, therefore they are much denser (almost two orders of magnitude) than the surrounding coronal plasma. Let us suppose that the spicules move along magnetic field lines of expanding flux tubes, which has much larger radius than the spicule itself (Fig. 3). The magnetic field strength will be decreased with height as well as in the radial direction. Consequently, a spicule moving along the field line away from the tube center will "feel" more decreased field strength than that moves along the field line at the tube center. The field components in current-free axisymmetric expanding magnetic flux tubes are
\begin{equation}\label{4}
B_r=B_0J_1 \left ({r}/{H_B}\right )e^{-z/H_B}, \\ B_z=B_0J_0 \left ({r}/{H_B}\right )e^{-z/H_B},
\end{equation}
where $H_B$ is a scale height of the magnetic field. The Alfv\'en speed in the transition region, say at $z=0$, can be assumed as 150 km s$^{-1}$ at the tube center, $r=0$. Pressure scale height for the transition region temperature of 0.1 MK is around 5 Mm, which gives $H_B$=10 Mm. 

We first consider a spicule, which start to flow along the field line at the tube center, $r=0$. At the height of $z=H_B/2=5$ Mm, the magnetic field will drop by $exp(1/2)=1.65$  and the Alfv\'en speed will become 90 km s$^{-1}$ (density decrease with height will have much less influence). Then the critical flow speed will be $90\cdot 1.25 \approx 115$ km s$^{-1}$ (using Eq. 3) and the jet moving along the tube center with the speed of 100 km s$^{-1}$ will by completely stable up to 5 Mm height, {\bf a}t least.

We now consider a spicule, which start to flow along the field line at the distance of $r=H_B/2$ from the tube center, near $z=0$. The magnetic field strength, $\sqrt{B^2_r+B^2_z}$, will be decreased in height as well as in the radial direction. At $z=0$, the field strength is decreased by 0.97 at the distance of $r=H_B/2$. In the simple case considered here, the plasma density is decreasing upwards owing to the stratification, while it stays almost constant in the horizontal direction. Therefore, the Alfv\'en speed will remain unchanged being 145 km s$^{-1}$. As the spicule follows the same magnetic field line, then it will appear on the distance of $r \approx 1.4 \, H_B$ at 5 Mm height. The Alfv\'en speed will be around 70 km s$^{-1}$ there, which means the critical flow speed of $70 \cdot 1.25 \approx 90$ km s$^{-1}$ and the spicule moving with 100 km s$^{-1}$ will become unstable at this height.

Using Eq. (4) one can find that the angle of magnetic field line with regards to the vertical at $z=0$ and $r=H_B/2$ is around $15^0$ degree. As spicules generally follow the magnetic field lines, the angle of spicule axis with regards to the vertical will be also 15 degree at this position.  Any spicule with the same initial speed (in our case 100 km s$^{-1}$), which are inclined with more than $15^0$ degree, may become super-Alfv\'enic with height and hence unstable. The spicules with smaller speed will require more initial inclination angle in order to be unstable with height. It is of remarkable importance that observed median inclination angle of spicules from the vertical is $20^0-40^0$ degree \citep{Beckers1968,Tsiropoula2012}. To our knowledge, there is no statistical study of inclination angle for type II spicules.

One can estimate the periods and growth times of the unstable harmonics in the conditions of spicules. Typical diameters of type I and II spicules can be assumed as 400 km and 100 km, respectively, while the typical flow speeds are 30 km s$^{-1}$ and 100 km s$^{-1}$, respectively. The type I spicules are sub-Alfv\'enic and hence stable, while type II spicules are close to the Alfv\'en speed and, perhaps, some may be in the instability region. The dependence of periods and growth times of the unstable harmonics with $k_z a=0.1$ on the Alfv\'en Mach number is shown on Fig. 4. The period of unstable harmonics in type II spicules is almost independent of flow parameters and is around 31-32 s, while the growth time crucially depends on Alfv\'en Mach number and the density ratio. For the Mach number of $V_{Ae}/U_z=0.6$, the growth time is 100 s for $\rho_0/\rho_e=100$ and 45 s for $\rho_0/\rho_e=25$. Hence, type II spicules are strongly unstable to the dynamic kink instability and the growth time has the same order as the period. It is interesting that the growth time is of the same order as the observed life time of type II spicules in Ca II H line, therefore one can suggest the connection between the instability and observed transverse motion of the structures. One can assume that the axes of type II spicules start rapid transverse motions owing to the instability and make back-forth swing when the growth time is longer than the period i. e. for $1.25 \, V_{Ae}<U_z<2 \, V_{Ae}$. On the other hand, the spicule may show only linear transverse motion without full swing when the growth time is shorter than the period i. e. for $U_z>2 \, V_{Ae}$. This means that the very high-speed spicules may be destroyed owing to the instability, before they will return back to the initial position.

 \section{Discussion and conclusion}

Observed transverse displacement of spicule axis has been suggested to be caused either by volume-feeling Alfv\'en waves or MHD kink waves \citep{Kukhianidze2006, DePontieu2007b,Zaqarashvili2009,Okamoto2011}. In both cases, the transverse motion is resulted by the Lorentz force acting on magnetic field lines. Here, we suggest an alternative mechanism for the transverse displacement. The mechanism is connected to the centripetal force of jet flowing along the curved trajectory: when the axis of hydrodynamic jet is deformed, then the force tries to amplify the displacement and leads to the kink instability as shown on Figure 1. If the jet flows along the magnetic field, then the Lorentz force tries to straighten the field lines and hence acting against instability. The ratio of Alfv\'en to the jet speeds defines the critical threshold for the dynamic kink instability. For the dense jets in rarified environment (resembling spicules) the instability may start for $U_z>1.25 \, V_{Ae}$. The velocity of type I spicules is less than Alfv\'en speed in the lower corona. But the speed of type II spicules is comparable (perhaps under certain circumstances) to the external Alfv\'en speed, which may lead to the dynamic kink instability in these structures. 

Spicules probably follow the magnetic field lines when they rise up into the corona. The magnetic field obviously changes its structure from the chromosphere to the corona expanding upwards, which suggests that the Alfv\'en speed should be also inhomogeneous in vertical and horizontal directions. Therefore, the conditions for the dynamic kink instability for type II spicules may arise only in certain regions where the flow become super-Alfv\'enic. This may happen in peripheries of expanding magnetic flux tube (Figure 3), where the negative gradient of Alfv\'en speed has maximal value owing to the volume expansion of magnetic field lines and minimisation of density stratification effect \citep{Hollweg1982,Grant2018}. Spicules flowing along the field lines at central axis of expanding magnetic flux tube may stay sub-Alfv\'enic and hence stable to the dynamic kink instability. On the other hand, spicules starting to flow along the field lines away from the tube axis (as on Figure 4) may become super-Alfv\'enic owing to the negative gradient of Alfv\'en speed and hence unstable. In the case of simple expanding magnetic flux tube with the Alfv\'en speed of 150 km s$^{-1}$ at $z=0$, the jet with 100 km s$^{-1}$ speed become unstable to the dynamic kink instability at 5 Mm height if it starts to flow along the field line at the distance $r=H_B/2$ from the tube center. 
This field line has the inclination angle of 15$^0$, therefore the spicule will be also inclined with the same angle to the vertical. All spicules started with $>$ 15$^0$ angle at $z=0$ (corresponding to the upper boundary of the chromosphere) with the speed of 100 km s$^{-1}$ can become unstable when they move upwards assuming the physical properties and constraints assumed here. More inclined spicules must show stronger transverse velocity. It may also happen that less inclined spicules may show full swinging transverse motion, while the more inclined ones only linear transverse trend. It could be interesting to check the dependence of transverse velocity and dynamics on inclination angle of type II spicules observationally.

The rapid disappearance of type II spicules in chromospheric spectral lines might be also connected to the dynamic kink instability as it may lead to the destroy of the structure. However, most of the spicules appear in hotter transition region lines \citep{Pereira2014,Rouppe2015,Skogsrud2015,Chintzoglou2018}, which probably indicate to their rapid heating  during transverse motion rather than decomposition. Heating mechanism might be ion-neutral collisions \citep{Erdelyi2004,Martinez2017}, Kelvin-Helmholtz instability \citep{Antolin2018} or both effects simultaneously \citep{Kuridze2016}.

Periods and growth times of unstable modes are estimated as $\sim$ 30 s and 25-100 s, respectively, in the conditions of type II spicules. On the other hand, the type I spicules seem to be generally stable to the kink instability.  It is interesting to note that the period and growth times of unstable harmonics in type II spicules are comparable to the range of their life times in chromospheric lines. This may indicate to the connection of dynamic kink instability to the evolution of these structures. The suggested scenario is simple: the type II spicules moving upwards with an angle to the vertical (for example, near the peripheries of expanding magnetic flux tube) may become super-Alfv\'enic at certain heights owing the negative gradient of Alfv\'en speed  and hence unstable to the dynamic kink instability. Then the axis of the spicules may start transverse motions, which might lead either to the complete destroy of the spicule owing to the instability or to the rapid heating by some mechanism up to the transition region temperature. In the first case, spicules will rapidly disappear in all spectral lines. In the second case, spicules will disappear in chromospheric spectral lines, but appear in hotter TR lines as it is found by observations. Inclination angle of type II spicules might play a significant role in both processes, which can be tested by observations.

It must be noted that consideration in this paper is rather simple catching only basic properties of instability and does not take into account observed velocity/density gradients \citep{Sekse2012} as well as torsional motions \citep{DePontieu2012}. Therefore, more analytical/numerical and observational study is necessary to perform in the future.

\acknowledgments

The work was funded by the Austrian Science Fund (FWF, project P30695-N27).

\appendix

\section{Derivation of dispersion equation governing a homogeneous jet in a magnetic field}

We consider homogeneous medium with uniform density, $\rho_e$, and homogeneous magnetic field, $B_z$, directed along the $z$ axis of cylindrical system $(r, \phi, z)$. Homogeneous cylindrical jet with radius $a$ and density $\rho_0$ moves along the magnetic field with the uniform velocity $U_z$. Using Fourier analysis with $\exp{(i m \phi +i k_z z -i \omega t)}$, where $\omega$ is the wave frequency and $m, k_z$ are wave numbers, one can readily find that the incompressible linear dynamics of perturbations is governed by the modified Bessel equation
\begin{equation}\label{5}
\frac{d^2 p_t}{dr^2}+\frac{1}{r}\frac{dp_t}{dr}-\left ( \frac{m^2}{r^2}+k^2_z\right )p_t=0,
\end{equation}
where $p_t=p+B_z b_z/4\pi$ in total (thermal + magnetic) pressure perturbation. Solutions for the total pressure inside ($r<a$) and outside ($r>a$) the jet are $p_{ti}=a_iI_m(k_z r)$ and $p_{te}=a_eK_m(k_z r)$, respectively, where $I_m$ and $K_m$ are the modified Bessel functions of order $m$ and $a_i, a_e$ are constants (note that the solution outside the jet is bounded at infinity). Then the Lagrangian radial displacement is governed by the expressions 
\begin{equation}\label{6}
\xi_{ri}=\frac{a_i}{\rho_i}\frac{ k_z I'_m(k_z r)}{(\omega - k_zU_z)^2-k^2_zV^2_{A0}}, \,\, \xi_{re}=\frac{a_e}{\rho_e}\frac{k_z K'_m(k_z r)}{\omega^2-k^2_zV^2_{Ae}},
\end{equation}
where $V_{Ae}=B_z/\sqrt{4\pi\rho_e}$ and $V_{A0}=B_z/\sqrt{4\pi\rho_0}$ are Alfv\'en speeds outside and inside the jet, while prime sign $'$ means the differentiation with Bessel function argument. Standard conditions of continuity of total pressure and displacement at the jet boundary ($p_{ti}(a)=p_{te}(a)$, $\xi_{ri}(a)=\xi_{re}(a)$) produces the dispersion equation
\begin{equation}\label{7}
\frac{\rho_e K_m(k_z a) I'_m(k_z a)}{(\omega - k_zU_z)^2-k^2_zV^2_{A0}}=\frac{\rho_i I_m(k_z a) K'_m(k_z a)}{\omega^2-k^2_zV^2_{Ae}}.
\end{equation}
From this equation one can readily obtain Eq. (1) of the main text
\begin{equation}\label{8}
(\rho_e E_m-\rho_0)\omega^2+2\rho_0k_z U_z \omega-(\rho_0k^2_z U^2_z+\rho_ek^2_z V^2_{Ae} E_m-\rho_0k^2_zV^2_{A0})=0,
\end{equation}
where  $E_m(k_z a)=(I'_m(k_z a)/I_m(k_z a))(K_m(k_z a)/K'_m(k_z a))$.

\bibliography{ms}{}
\bibliographystyle{aasjournal}



\begin{figure}[ht!]
\plotone{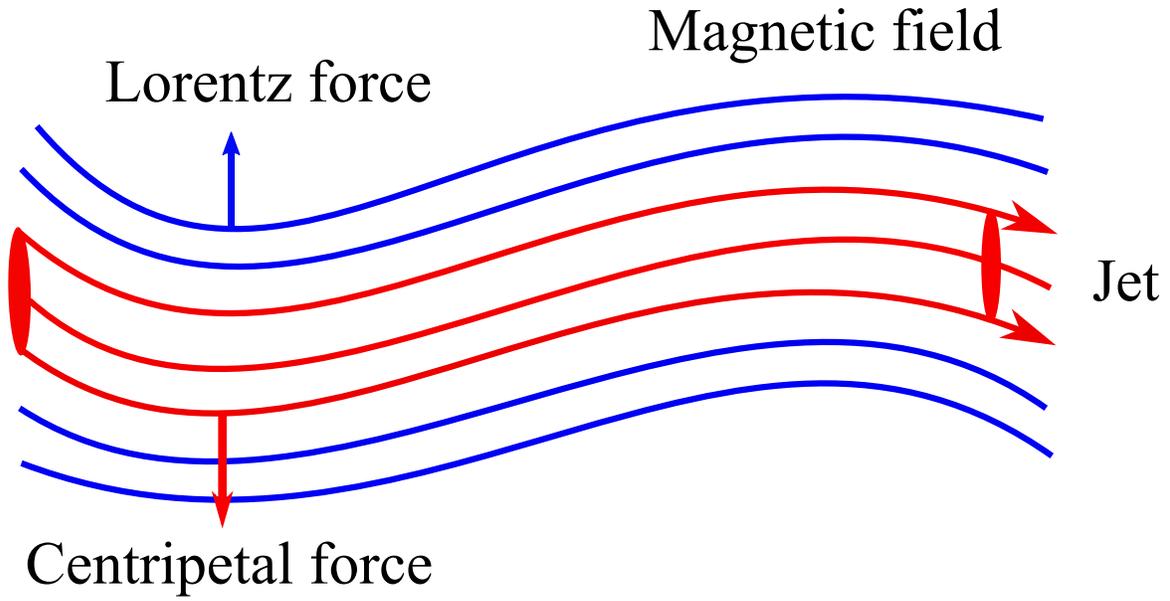}
\caption{Dynamic kink instability of jets (red lines) in the magnetic field (blue lines). {\bf Red (blue) arrow shows the direction of the centripetal force of flow (Lorentz force).} The centripetal force tries to amplify the transverse displacement, while the Lorentz force tries to reduce it. When the centripetal force is stronger then the displacement is unstable. \label{fig:general}}
\end{figure}

\begin{figure}[ht!]
\plotone{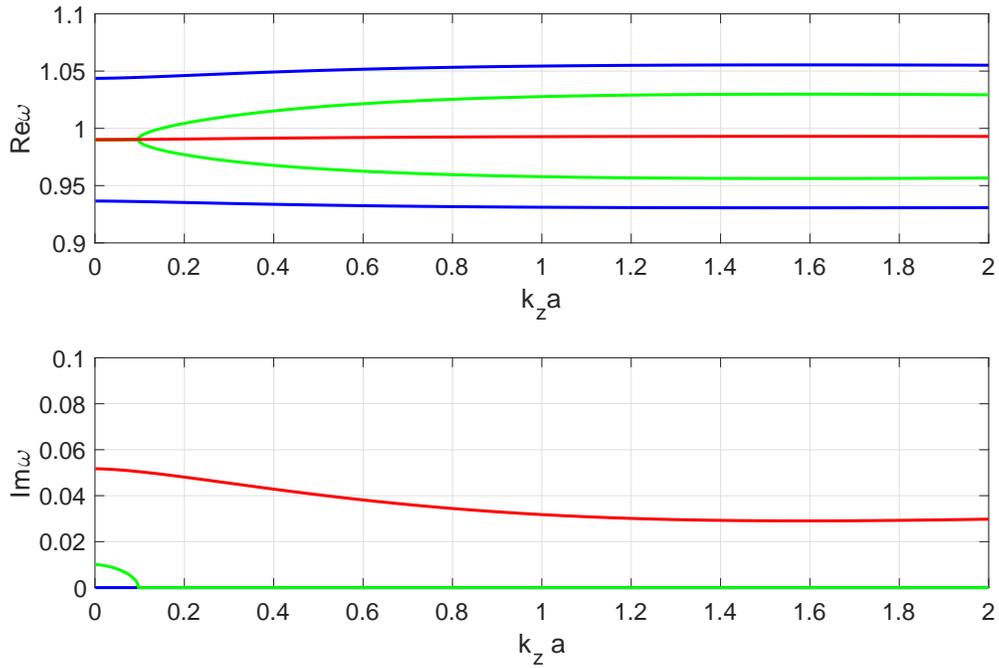}
\caption{Real and imaginary parts of frequency vs longitudinal wave number, $k_z a$. Blue, green and red lines correspond to $V_{Ae}/U_z$=0.8, 0.7 and 0.6, respectively. 
Here $\rho_0/\rho_e=100$ and the frequency is normalised by $k_z U_z$.
  \label{fig:general}}
\end{figure}

\begin{figure}[ht!]
\plotone{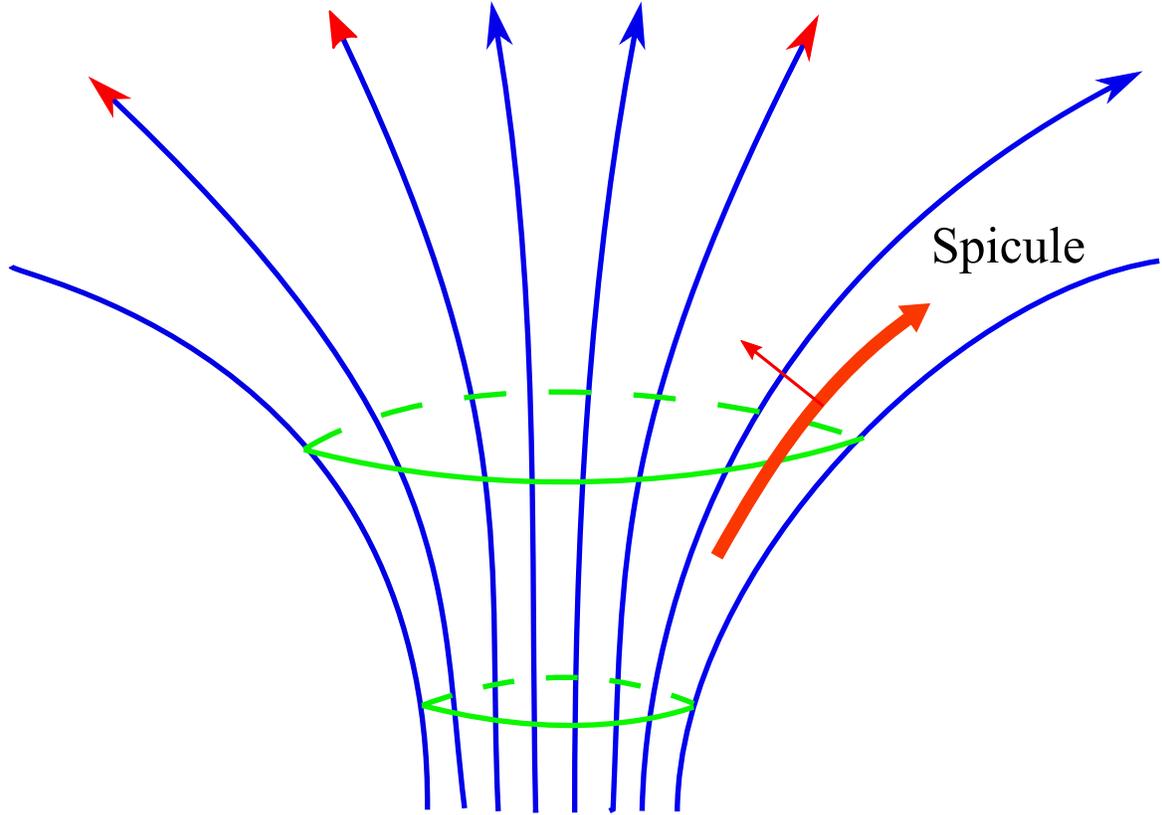}
\caption{The sketch of expanding magnetic flux tube. The lower green circle corresponds to the upper boundary of the chromosphere, while the upper circle may correspond to the height of $z=H_B/2$. The jet resembling spicule (thick red arrow) moves along magnetic field lines with an angle to the vertical. Thin red arrow shows the direction of centripetal force when the jet moves along a curved trajectory.   \label{fig:general}}
\end{figure}

\begin{figure}[ht!]
\plotone{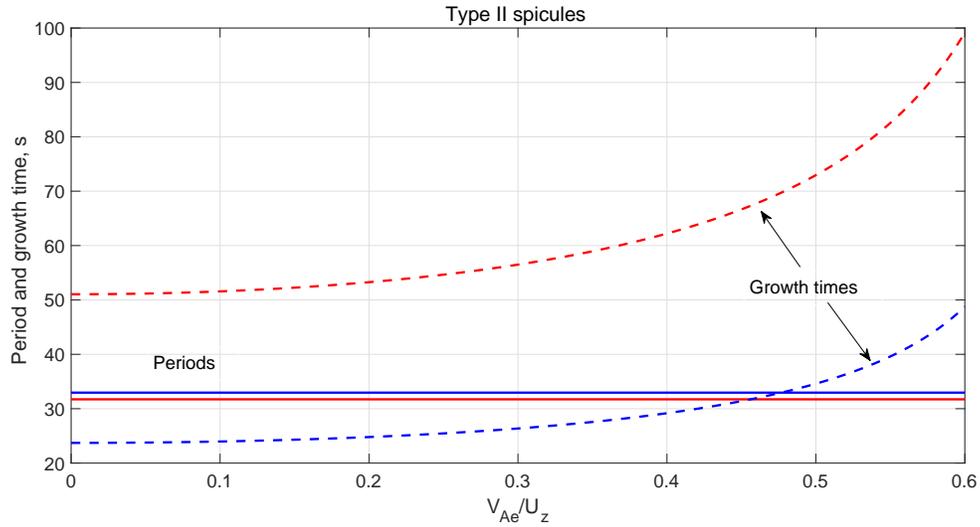}
\caption{Periods (solid lines) and growth times (dashed lines) of unstable modes with $k_z a=0.1$ vs Alfv\'en Mach number $V_{Ae}/U_z$. The flow speed and diameter of type II spicules are assumed to be $U_z$= 100 km s$^{-1}$ and $2 a=$ 100 km, respectively. Blue and red colours correspond to the density difference of $\rho_0/\rho_e=25$ and $\rho_0/\rho_e=100$, respectively. 
  \label{fig:general}}
\end{figure}


\end{document}